# Income, health, and spurious cointegration

José A. Tapia Granados[1] and Edward L. Ionides[2]

[1]Department of Politics and Program in Philosophy, Politics & Economics, Drexel University, Philadelphia, PA (jat368@drexel.edu).
[2]Department of Statistics, University of Michigan, Ann Arbor, MI (ionides@umich.edu).

*Abstract* –Data for many nations show a long-run increase, over many decades, of income, indexed by GDP per capita, and population health, indexed by mortality or life expectancy at birth (LEB). However, the short-run and long-run relationships between these variables have been interpreted in different ways, and many controversies remain open. It has been claimed that population health and income are cointegrated, and that this demonstrates a positive long-run effect of income on population health. We show, however, that an empirically tested cointegration between LEB and GDP per capita is not a sound method to infer a causal link. For a given country it is easy to find computer-generated data or time series of real observations, related or unrelated to the country, that according to standard methods, are also cointegrated with the country's LEB. More generally, given a trending time series, it is easy to find other series, observational or artificial, that appear cointegrated with it. Thus, standard cointegration methodology, often used in empirical investigations, cannot distinguish whether cointegration relationships are spurious or causal.

## 1. Introduction

This article investigates cointegration as a tool for causal reasoning in the social sciences via a case study of the link between life expectancy at birth (LEB) and gross domestic product (GDP). In Section 2, we describe the health economics debate, explaining the controversies surrounding the effects on population health of fluctuations in GDP. In Section 3, we introduce the cointegration approaches used to address this debate. In Section 4, we survey the applications of cointegration in this context, focusing in detail on two case studies. In Section 5, we investigate consequences of following the standard cointegration methods demonstrated in these case studies. The discussion in Section 6 places these results in the context of previous concerns about cointegration methodology.

## 2. The secular improvement of health and its disputed causes

In ancient Greece, Hippocrates claimed the effect of winds, waters, and types of housing on the health of city inhabitants, which would be also influenced by their lifestyle—whether they were or not drinkers and gluttons, or athletics and active (Buck et al. 1988). More reliable conjectures based on quantitative data emerged in the mid-17th century, when John Graunt and William Petty speculated on the health of the people using the weekly bills of mortality in London (Hull 1899). Graunt and Petty speculated that perhaps Londoners died at high rates because of crowding and continuous exposure to the smoke of coal and wood. Only continuous arrivals of people from the countryside allowed the city to grow because, annually, there were more deaths than births (Cairns 1997).

One century and a half later, in 1798, Malthus published his tract *On Population*, in which he posed a mutual interaction between population and income. Malthus claimed that a higher income would stimulate population growth which first would reduce available resources per person which in turn would trigger famines and wars that would restore the equilibrium by reducing population to a level consistent with available resources (Malthus 1970; Fogel 1994). These Malthusian speculations were largely discredited when statistics on deaths, population and income started to be available at different times in each country. Data going back for some countries even to the 18th century showed that, in the long-run, both health progress—as measured by declining mortality rates and the correspondent rising life expectancy at all ages— and income growth—as proxied by increasing GDP per capita—had occurred.

Life expectancy at birth (LEB, $e_0$ in demographic notation) is a major indicator of population health that uses all the age-specific mortality rates observed in a single year and assumes that a hypothetical cohort would be exposed to those rates until death of all its members; LEB is the mean length of life of the individuals in that hypothetical cohort. This is *period* LEB; the average lifespan of individuals born in a particular year is called *cohort* LEB and is not relevant in this context (ONS 2021). In the long run, LEB has increased in each country, and years of decrease in LEB can be characterized as health crises (Riley 2001). These health crises were common in Europe before the 20th century (Figure 1), associated with epidemics, generally triggered by poor harvests that lead to famine (Tapia Granados & Ionides 2008). They were associated with major declines in income in the predominantly agricultural economies of the time. Since the start of the 20th century, crises of high mortality with associated drops in LEB were mostly observed in exceptional periods of pandemic, war, famine, or major sociopolitical crisis. Thus, major drops in LEB occurred worldwide in 1918, during the world flu pandemic (Crosby 2003). More localized health crises with reductions in LEB occurred in the USSR during the mass repression and famine of the 1930s (Haynes & Husan 2003); during World War II in most European countries, as well as in Indian Bengal and the Chinese province of Henan; and during the early 1990s in the countries of the former Soviet bloc (Sen 1981; Cockerham 1999; Stillman 2006; Mackenbach 2013; Tapia Granados 2013, 2022). Figure 1 illustrates the long-term evolution of LEB in three European countries, showing the health crisis connected with the epidemics of infectious diseases generally associated to famines in pre 20th century Sweden, the world flu pandemic in 1917, and the effects of World War II in England and Wales and in the Netherlands, where the German occupation caused a major famine in the winter of 1944-1945. Figure 2 shows the remarkable evolution of LEB in several countries since the 1990s, a decade that in terms of population health was a major disaster for the countries that had been part of the USSR. The impact of the COVID-19 pandemic in 2020 is obvious in the drop in LEB shown in almost all countries in the figure, though, in the cases of the United States and Cuba, LEB was already declining from the mid-2010s.

Income decline in market economies is characteristic of periods of economic crisis, that is, recessions or depressions, like the Great Depression of the early 1930s, the Great Recession of 2009, or the COVID-19 depression of 2020. In 2009, during the Great Recession, the rate of growth of GDP per capita was -3.7% in the United States and -11.4% in the United Kingdom, while in the pandemic year 2020, it was -3.5% in the US and -5.2% in the UK (WDI 2023). In 2020-2021 COVID-19 caused a global mortality estimated in 18 million deaths (COVID-19 Excess Mortality Collaborators 2022) with the consequent drop in LEB in every nation. In the US, LEB declined from 78.8 in 2019 to 77.0 in 2020 (WDI 2023). The recession of 2020 was thus associated with a quite dramatic worsening of population health. However, that was a very special case because both in the Great Recession of 2009 and the Great Depression of 1930-1933, falling incomes were associated with decreasing mortality and rising LEB (Tapia Granados & Diez Roux 2009; Tapia Granados & Ionides 2017; Finkelstein et al. 2024). Indeed, many investigations since the 1970s (Eyer 1977a; Ruhm 2000, 2005; Tapia Granados 2005; Gerdtham & Ruhm 2006) have supported early finding by Dorothy Thomas in the 1920s of a procyclical oscillation of death rates, so that over and above its declining long-term trend, mortality rises in business cycle expansions and falls in recessions (Ogburn & Thomas 1922; Thomas 1927).

Of course, many of these notions are controversial. To describe the evolution of time series is easier than inferring causal links between them.

Attempts to explain the long-run fall of death rates "did not begin until after World War I because before that time it was uncertain whether such a decline was in progress" (Fogel 1993). Until the 20th century, Malthusian ideas and the miasmatic theory—that posed as causes of disease foul odors, *miasma,* produced by filth, decay, and putrefaction—were probably the vague notions underlying the prevalent idea that health problems were somewhat connected with poverty. Thus, in his *Principles of Economics*, Alfred Marshall stated that "physical, mental, and moral ill-health is partly due to other causes than poverty: but this is the chief cause" (Marshall 1920).

During most of the 20th century, it was a common view that improvements in sanitation and personal hygiene as well as biomedical advances for treating specific diseases were together with the reduction of poverty the major factors explaining the secular decline of mortality (Fogel 1991, 1997). It was for that reason that the 1920s observations by Dorothy Thomas of rising mortality in periods of economic prosperity were puzzling for the few who paid attention to them. Major puzzlement emerged again when rather than increasing, mortality rates *declined* in the United States during the Great Depression (Sydenstricker 1934; Wiehl 1935).

The "demographic transition"—the long-term evolution of societies from high to low levels of fertility and mortality—and the "epidemiological transition"—the displacement of infectious diseases by chronic and non-communicable diseases as major causes of death—were often interpreted in the mid decades of the 20$^{th}$ century as revealing the effects of antibiotics and other therapeutical innovations. However, in the 1970s a medical historian, Thomas McKeown, proposed an alternative explanation. By showing that the decline of death rates due to infectious diseases had started much before than specific means to treat these diseases were available, McKeown questioned the idea that biomedical advances had contributed decisively to reduce mortality. He claimed that generalized improvements of nutrition had occurred associated with the increasing availability of food in an economic environment of rising income, which had led to improved immune resistance to infectious diseases; this would be the key factor explaining the secular decline of mortality rates since the 19th century (McKeown 1976, 1988). Arguing against McKeown, the demographer Simon Szreter (1988, 1998, 1999) emphasized the role of a variety of public health-enhancing processes—such as the implementation of water supply and sewage infrastructure, improvements in housing, pasteurization of milk, and widespread vaccination—in the long-run decline of mortality.

In 1993 Robert W. Fogel received the Nobel Memorial Prize in Economics for his research in economic and demographic history. In his research, Fogel highlighted the influence of poor caloric intake to explain the small body size and the short life span until the 18$^{th}$ century, after which nutrition started to improve leading to decreasing the incidence of disease, raising height, and increasing labor productivity, in a network of effects interlaced through multiple synergies. Perhaps unknown to him, Fogel did not cite Thomas's, McKeown's, or Szreter's contributions in his Nobel lecture. He asserted, however, that "for many European nations before the middle of the 19th century, the national production of food was at such low levels that the poorer classes were bound to have been malnourished under any conceivable circumstance", so that "the high disease rates of the period were not merely a cause of malnutrition but undoubtedly, to a considerable degree, a consequence of exceedingly poor diets." This was an idea that McKeown had emphasized in the 1970s.

The period of very low unemployment rates in Western countries following World War II ended in the 1970s, and it was at that time when controversies on the way business cycles may have an effect on mortality occurred for the first time. In a path-breaking paper that cited studies showing that mortality had decreased during the Great Depression, Joseph Eyer showed data that plainly demonstrated how in the US mortality had oscillated procyclically for a century, i.e. upward in expansions and downward in recessions (Eyer 1977a). He attributed such oscillation to social stress (e.g., overtime, increased transportation, pollution and consumption of harmful substances) enhanced by the acceleration of economic activity during business cycle expansions (Eyer 1977b). Interestingly, in historical demography something similar had been observed. When real wages had been high in England during periods of the 16th and 17th century, there had been increases in mortality: "It is reasonable to suppose that in certain circumstances improving living standards will tend to raise mortality rather than reduce it. If higher real wages (…) concentrate more people in cities, higher death rates may result" (Wrigley & Schofield 1981:415).

Harvey Brenner was an enthusiast of economic growth and a deft propagator of his views.

In contributions expanding four decades and including a report to a committee of the US Congress, Brenner (1971, 1984, 2005, 2011) claimed that economic growth was the key cause for mortality decline, with recessions having a major harmful effect on mortality. For Brenner the increase in heart attacks or deaths because of respiratory disease that was observed during expansions was *a lagged effect* of the previous recession. Brenner was never specific about how long the lag was, and never addressed the notion that, for instance, an increase of heart attacks or respiratory deaths in the expansion of the business cycle is fully consistent with the fact that smoking, overtime, and industrial pollution are procyclical (Mitchell 1951:71).

Christopher Ruhm's 2001 paper "Are recessions good for your health?" was a keystone in introducing the notion of procyclical mortality in the economic literature. For a while, it looked like Ruhm's contribution had settled the questions on the procyclical oscillation of mortality, but the present century saw a continuation of controversies that somewhat reenacted the Eyer-Brenner exchanges of the 1970s (Brenner 2005; Catalano & Bellows 2005; Edwards 2005; McKee & Suhrcke 2005; Neumayer 2005; Ruhm 2005; Tapia Granados 2005a, 2005b, 2013; Stuckler & Basu 2013; Catalano & Bruckner 2016; Tapia Granados & Ionides 2016). Overall, however, the notion that, after exclusion of long-term trends, mortality is procyclical, has been subjected to multiple replications. For example, a recent examination of the impact of the Great Recession on mortality in the United States by Finkelstein et al. (2024) concluded that during this recession mortality reductions occurred across causes of death and were concentrated in the population with a low level of education, with declines in elderly mortality explaining about three-quarters of the total mortality reduction during the recession. Finkelstein et al. concluded that recession-induced mortality declines depend primarily on external effects of a decrease in business activity on mortality, in which recession-induced drops in air pollution would be a key mechanism. This study, largely consistent with previous work by Eyer (1977a), Kasl (1979), Ruhm (2005), Gerdham & Ruhm (2006), Tapia Granados & Ionides (2008, 2016, 2017), and many others (e.g., Lindo 2015; Haaland & Telle 2015; van den Berg et al. 2017) is further evidence against Brenner's notions.

Considerations about how economic distress may affect mortality differently according to the level of economic development led to Gonzalez & Quast (2010a; 2010b) to analyze panel mortality data from the Mexican states. They found that state mortality had oscillated procyclically or countercyclically depending on the level of economic development of the state. Other studies in low- or middle-income countries found, however, just a procyclical oscillation of mortality (Lin 2009; Leveau et al. 2021). But the notion that recession must be harmful for population health apparently does not want to die, and reemerges supported by faulty reasoning. A recent paper by Doerr and Hoffman (2023) uses panel data covering 180 countries over six decades to reach the conclusion that recessions are systematically associated with higher mortality rates, a result that is explained because the years of social chaos in the early 1990s in the countries of the old Soviet bloc, when death rates skyrocketed, are considered in that paper as "recessions".

Versus the common emphasis of Brenner and many others in economic growth as the basis of mortality decline, the demographer Samuel Preston found in the 1970s that only about a fifth of the decrease in mortality rates between the 1900s and the 1960s could be attributed to increases in material affluence (GDP per capita). These studies by Preston (1975, 1976) were cited thousand times and set a standard of rigor in demographic literature. Three decades later, taking his previous beliefs further in the direction of questioning the link between income growth and health progress, Preston explained that modern investigations on the years 1960-2000 showed almost no relation between changes in LEB and economic growth (Preston 2007). Indeed, remarkable declines in mortality with little or no economic growth had occurred in many countries, and in India and China a negative correlation between decadal rates of GDP growth and reductions in child mortality was observed (Cutler, Deaton & Lleras-Muney 2006; Preston 2007).

Also from the field of demography, Richard Easterlin questioned the notion that the decline of mortality in the 20th century had been simply an effect of economic growth. If that were the case, it would be difficult to explain the delayed start of the mortality decline and its more rapid spread, as well as the international convergence in LEB since the mid-20th century, despite big differences in growth of income per capita (Easterlin 2004:87; 1999).

Authors like McKeown, Szreter, Fogel and others might be widely interpreted as suggesting a causal long-term effect of economic development or, in other words, rising levels of income, on population health. Improved nutrition, generalized vaccinations, cleaner water and more hygienic milk, and other factors that can be considered part of "development" would in a synergic way lead to higher immune resistance to infections. Hygienic measures and behaviors and sanitary infrastructures would reduce transmission of germs. All this together with better housing, better conditions of work and better medical treatments for diseases, would lead to declining mortality. Note however that many of these potential mechanisms to explain the decline in death rates refer to processes that occurred once in historical time and are not cumulative. Thus, the improvement in nutrition associated with elimination of deficient caloric intake or intake of specific nutrients can be linked to increasing income, but once it has been accomplished, additional increase in income may not lead to further improvement in nutrition. Indeed, it can lead to excessive and harmful caloric intake—as clearly illustrated by the modern worldwide epidemics of obesity and its associated chronic diseases such as diabetes and cardiovascular disorders. Similarly, the supply of clean water, the elimination of sources of infectious germs by development of urban sanitary infrastructures, and the generalized vaccination of children is linked to income growth, but once all that is in place, additional income growth will not lead to further improvements in these health-promoting factors. These are just some examples of the many potential mechanisms proposed to explain the complex relationship between income and population health.

In econometric studies that generally have ignored Preston's contributions as well as the literature demonstrating a procyclical oscillation of mortality, statistical tests for cointegration have been used to infer causal relationships between mortality and income. Both the level of population health—as measured by rising LEB or declining mortality—and income—as measured by GDP per capita—are trending variables and some authors have claimed that they are cointegrated and thus causally linked. In the following sections we present first some considerations on cointegration; second, some statistical results that show how the notion of cointegration is not useful to analyze the relationship between LEB and GDP per capita in the long run; third, we report the results of exploring cointegration between a variety of time-series, finding evidence for spurious cointegration in multiple cases. In the concluding section we discuss the relevance of some general ideas on the notions of stationarity and cointegration, and the use and misuse of statistical tests concerning them.

## 3. Cointegration

Cointegration is defined as existing "when two or more time series variables share a common stochastic trend" (Stock and Watson 2019: 734). But, what is a stochastic trend?

In technical terms, a time series model has trends when it is not stationary, and it is stationary "if it has time invariant first and second moments" (Lütkepohl 2005: 235), meaning its mean, standard deviation, and lagged correlations are constant through time. Time series such as population size, GDP per capita, female share in the labor force, infant mortality, or the illiteracy rate show obvious rising or falling trends. The sample mean of this type of time series varies depending on the period considered, so these series should unambiguously be modeled as non-stationary. A trend can be modeled as a deterministic function (say, linear or exponential) or as a stochastic process. However, as stated in an important textbook, econometricians think "it is more appropriate to model economic time series as having stochastic rather than deterministic trends" (Stock and Watson 2019:541). There is therefore an important predisposition in econometrics *to consider that any trend is a stochastic one*.

When a non-stationary series, $x_t$, is transformed by converting it to first differences ($\Delta x_t = x_t - x_{t-1}$) or rate of change ($[x_t - x_{t-1}]/x_{t-1}$, which is approximately equal to the logarithmic difference ($[x_t - x_{t-1}]/x_{t-1} \approx \ln x_t - \ln x_{t-1}$), the result is often a series that looks in a graph as a stationary oscillation around a mean value. For mathematical reasons that are not to be discussed here, in econometrics it is often said that a series that is non-stationary has a "unit root," which has the technical meaning that the first difference is a stationary linear stochastic process. However, it is worth bearing in mind that not all non-stationary time series models have a unit root. For example, a series with an exponential trend plus a random Gaussian error remains non-stationary after any amount of differencing, and does not possess a unit root.

If a time series becomes stationary when differenced, the original series is called integrated of order one, or I(1), while the series resulting from differencing is called integrated of order zero, I(0). Thus, by definition, if $x_t \sim$ I(1), with ~ meaning "is", then $\Delta x_t \sim$ I(0), as stated by Engle & Granger (1987) in one of the early canonical papers on cointegration.

Strictly speaking, common assertions such as "LEB is I(1)" or "GDP is I(1)" or "the rate of growth of GDP is I(0)" are inexact. The notion of integration of order 1, I(1) in econometric vernacular, applies to random processes, made up of random variables which are mathematical abstractions. But neither GDP nor LEB are random variables in this formal sense; contrarily, they are numbers derived from measurements of our world. We can consider modeling GDP, or LEB, or other time series pertaining to any social or natural science in many ways, some of which maybe I(1), or I(0), models. A particular series of annual values of GDP can be approximated with a linear or a polynomial equation, or an exponential trend, or a random walk with drift. But GDP is neither a value produced by a polynomial equation, nor a random walk, nor an exponential curve.

In this discussion, we follow the common abuse of notation in which we allow ourselves to talk about data being stationary or I(1), despite the fact that these are actually properties of models, not data. We do so because this widespread minor technical error is not the core problem with the use of cointegration in the studies we consider. Nevertheless, obscuring the distinction between models and data does have risks. For example, models can be misspecified, or more than one model can be plausible. Whenever we say that a time series is I(1), or a pair of time series are cointegrated, we invite the reader to recall that we mean only that a specific statistical test was carried out which indicated that the corresponding model is statistically plausible according to that particular test.

In statistics or econometrics courses, students are advised to be vigilant against the phenomenon of spurious regression, which may appear if the series are not stationary (Gujarati 2009:747). If we regress the annual data of the monetary aggregate M2 for the Spanish economy on the value of annual imports of Italy for the years 1941-1998, we obtain an effect estimate of 273.3 with a standard error of 3.7 which is highly significant ($P < 0.0001$). The regression $R^2$ is 0.990, so that the value of Italian imports in the years 1941-1998, measured in liras, explains 99% of the variation of the Spanish monetary base measured in pesetas. Of course, all this is nonsense because this is a spurious regression. That is because both series have a rising trend.

In 1974 Clive Granger and Paul Newbold raised concerns about spurious regressions that they saw as being very commonly reported in applied econometric literature. They explained that many economic time series in 'levels' have trends and "considerable care has to be taken in specifying one's equations" to avoid spurious regressions. They recommended "taking first differences of all variables that appear to be highly autocorrelated" which "should considerably improve the interpretability of the coefficients." However, economists found this procedure frustrating, as it was seen as a waste of potentially useful information on the causal relationships among series in their long-run evolution. Thus, the notion of cointegration probably was born from this frustration when it was introduced by the same Clive Granger in 1981 to indicate "a genuine relation" (Hendry & Juselius 2000:16) between two trending variables; cointegration was intended to be, then, the obverse of nonsense or spurious regression.

In the past four decades, economic literature has discussed cointegration between such time series like consumption and income, wages and prices, interest rates and inflation, prices of the same commodity in different locations, forward and spot exchange rates, inflation rates and interest rates, etc. (Hjalmarsson & Österholm 2007). The usual notion is that in the long-run, two cointegrated variables "evolve together," so that they do not "drift apart" too much. But the common notion is that cointegration between two variables indicates a genuine relation between them (Hendry & Juselius 2000), so that causality is implied. As it will be seen, this is explicit in the studies claiming cointegration of income and health.

The simplest case of cointegration of $x_t$ and $y_t$ appears when the difference $x_t - y_t$ is a constant quantity or, in a less simplistic way, the difference between the two variables, $x_t - y_t = z_t$ is a stationary variable. For example, data may show that, on average, the price of a product in urban area A is 63 cents lower than the price in region B. It could be also that $y_t$ is, let's say, an "augmented or diminished version" of $x_t$, say $\beta \cdot x_t$, in which case $y_t - \beta \cdot x_t = z_t$ is constant or a stationary series, that is, in econometric parlance, I(0). As Hendry and Juselius (2000) put it, for $\beta = 1$, "the vague idea that $x_t$ and $y_t$ cannot drift too far apart has been translated into the more precise statement that 'their difference will be I(0).' The use of the constant [$\beta$] merely suggests that some scaling needs to be used before the I(0) difference can be achieved."

Economic theory is often expressed in equilibrium terms. Equilibrium relationships are theorized between such variables as supply and demand, household income and expenditure, prices in different markets, or particular monetary aggregates. The notion of equilibrium is indeed customary in discussions on cointegration. For instance, gasoline 87 octane prices, observed daily in different locations of the same country should be "in equilibrium". If the equilibrium concept "is to have any relevance for the specification of econometric models, the economy should appear to prefer a small value of $z_t$ [the cointegrated series] rather than a large value" (Hendry & Juselius 2000). Within this theoretical structure, cointegration would be of considerable interest, since by determining stationary relations that hold between variables which are individually non-stationary, the cointegration relationship is useful to show "long-run equilibria" which act as "attractors", "towards which convergence occurs whenever there are departures therefrom" (Hendry & Juselius 2001:76). Thus, a classical textbook explains the cointegration of two variables as meaning that they are both I(1) and have a long-term, or equilibrium, relationship between them (Gujarati 2009:762).

Cointegration is intimately linked with the so-called ECMs, error-correction models, meaning that changes in a variable depend on the deviations from some equilibrium relation between them (Lütkepohl 2005:247). Indeed, econometricians have shown that cointegration and ECMs are "actually two names for the same thing: cointegration entails negative feedback involving the lagged levels of the variables, and a lagged feedback entails cointegration" (Hendry & Juselius 2000).

The usual procedure to test cointegration between two series starts by testing their stationarity. The augmented Dickey-Fuller (ADF) and the Phillips-Perron (PP) tests yield as test statistics a rho, and a tau (the ADF, also an F value), each one with an associated *P*-value. The tau value is usually the accepted test statistic. Each of these two tests has three different varieties (zero mean, single mean, and trend), and it should be tuned to the characteristics of the particular time series at hand, as a decision is needed on what are the proper autoregressive (AR) and moving average (MA) orders of the model on which the test is based. This is sometimes done by minimizing an information criterion such as AIC or BIC, though in the literature often just a small value or values for AR and MA are used. All these intricacies make the interpretation of the stationarity tests far from being straightforward. At any rate, if two series are considered I(1), the Johansen's rank cointegration test can be applied to them. It also requires choosing the appropriate AR and MA orders. The test renders a trace and a corresponding *P*-value. The null hypothesis $H_0$ is that the series are not cointegrated, thus small *P*-values (0.05 is the 95% confidence level) means rejection of $H_0$, that is evidence in favor of the alternative, that the series are cointegrated. Let's see now how all this applies to the case of GDP per capita and LEB.

## 4. Cointegration between income and health

In the long run mortality rates have declined while both LEB and GDP per capita have increased in all countries, and some authors have published claims of cointegration between these variables. This would indicate a causal link either from income to health, or from health to income, or in both directions. By contrast, in major studies on the determinants of mortality (Preston 1975, 1976, 2007; Easterlin 1999, ,2004; Cutler and Lleras-Muney 2006), cointegration has not been mentioned. Furthermore, studies by Acemoglu and Johnson (2007, 2014) have concluded, without using cointegration, that there is no observable effect of LEB on GDP growth.

To our knowledge, Suchit Arora (2001, 2005) was the first who claimed cointegration between GDP per capita and LEB. For Arora this relationship of cointegration in ten industrialized countries indicated that improvements in health increased the pace of economic growth by 30% to 40%.

Harvey Brenner posed an ECM, i.e., a cointegration model, linking income growth and declining mortality throughout the 20th century in the United States (Brenner 2005). He claimed this model demonstrated both the harmful effect of recessions on mortality and how 20th century improvements in health in the United States had its base in economic growth.

Robyn Swift (2011) claimed to have found cointegration between LEB and both total GDP and GDP per capita for 13 OECD countries over periods of many decades. Using Johansen's cointegration method, Swift concluded that a 1% increase in LEB results in an average 5% increase in GDP per capita in the long run, and in the other direction of causation, total GDP and GDP per capita also cause a significant effect on LEB.

The results by Brenner and Arora were contested. Specific criticisms were stated against the way Brenner depicted the possible links between mortality and the economy and Arora's results on cointegration of income and LEB in Britain were disputed on the basis of reproducibility (Tapia Granados 2005, 2012). Neither Brenner nor Arora replied to these criticisms. However, despite being disputed, Brenner's studies have been cited as if they had proved a cointegration link, that is, a causal link, between income and population health (Boonen & Li 2017; Niu & Melenberg 2014).

In an investigation published five years ago, Maddalena Cavichioli and Barbara Pistoresi used Johansen's test for cointegration applied to Italian annual data 1862-2013. They found general mortality and several of its components cointegrated with GDP per capita. They claimed to have avoided omitted variable bias or spurious associations, and interpreted the cointegration they found as demonstrating a causal link between income and health, so that "an increase of 1% in the real GDP per capita induces a reduction in mortality rate of 0.27% for the total population and 0.24% for the male group" (Cavichioli & Pistoresi, 2020).

More recently, a controversy has emerged on the cointegration of income and LEB in the United Kingdom, that was first denied by Tapia Granados (2012). This author explored the evolution of health and income in England and Wales using annual data 1840-2000 of LEB for England and Wales and GDP or GDP per capita for the UK. Tapia Granados (2012) compared the annual growth of LEB with the annual rate of growth of GDP per capita, finding a negative relation between the growth of both variables, so that the lower the rate of growth of the economy, the greater the annual increase in LEB (either total, male or female LEB). As supportive of his conclusions, Tapia Granados (2012) cited a study by Amartya Sen, who similarly had noticed that significant increases in LEB in England and Wales in 1901-1960 had occurred during decades of slow economic growth, so that a negative correlation appeared between decadal economic growth and decadal gains in LEB (Sen 2001). Overall, Tapia Granados (2012) interpreted his results as further evidence adding to an emerging consensus that in the context of long-term declining death rates, mortality is procyclical and LEB is countercyclical, as the decrease in mortality and the increase in LEB tend to be faster when economic growth slows down in the contraction phase of the business cycle. With a quite unsophisticated analysis (see below), the suitability of cointegration models between income, proxied by GDP per capita, and health, indexed by LEB, was dismissed by Tapia Granados (2012).

The controversy emerged when two years ago Chowdhury, Cook & Watson (2023, hereafter Chowdhury et al. 2023, revisited this issue and proved the cointegration of these series, explicitly rejecting Tapia Granados's conclusion of no cointegration. Chowdhury et al. showed that log-transformed, LEB of England and Wales, and GDP per capita of the UK can be modeled as non-stationary, I(1) series, which they showed are cointegrated. They asserted that the main outcome of the analyses they present "is the reversal of Tapia Granados's conclusion of no long-run relationship between life expectancy and income" in the specific case of England and Wales.

Chowdhury et al. used LEB data from the Human Mortality Database. From their graphs, it appears they analyze LEB for the civilian population, as used by Tapia Granados, and we do the same here.

Plotted without any transformation (Figure 3), the series of LEB and GDP per capita used by Chowdhury et al. (2023) and by Tapia Granados (2012), look quite unlikely to be "moving together", as the LEB series looks like a curve with upward convexity, that is, with a positive but declining slope, while GDP per capita shows an upward concavity, that is, a positive but increasing slope. This "graphical inconsistency" does not suggest cointegration and was one of the reasons leading Tapia Granados (2012) to rule out this possibility.

By transforming both series into natural logs (Figure 4 here, Figure 5 in Chowdhury et al. 2023) the two series become more linear and, using a suitable scale, they can be plotted to follow each other quite closely (Figure 4, upper panel). However, just by changing the scale, the log-transformed series can be also plotted as widely diverging (Figure 3, lower panel).

Some of the analysis by Chowdhury et al. involve unusual tests of cointegration that we were unable to replicate. Much of the paper focuses on investigating whether the natural logs of the series, ln GDP per capita and ln LEB, are or are not stationary, and whether there is a "break in trend" in the series, a break that they identify in 1917, the year of the world flu pandemic. For us it is obvious that LEB in England and Wales is a non-stationary series that departs from trend in 1914-1918 and 1940-1945, the years of the world wars (see Figures 1, 3, and 4). The years 1917-1918 of the world flu pandemic clearly belong to the general departure of trend that England and Wales's LEB had during World War I. That both the GDP per capita of the United Kingdom and LEB of England and Wales have a rising trend is obvious. Using the 1841-1999 data, for ln GDP per capita, both the augmented Dickey-Fuller (ADF) test and the Phillips-Perron (PP) test in their three varieties and with autoregressive (AR) parameter 1 or 2 yield $P$-values over 0.98 so that the null hypothesis of unit roots for ln GDP per capita cannot be rejected, and following the usual practice in hypothesis testing, the alternative shall be accepted; the conclusion is that ln GDP per capita has a unit root. However, for ln LEB the trend type of both the ADF and the PP tests yield values well below 0.05, so that the unit root null hypothesis has to be rejected; the series appears to be stationary. Following the notion of a potential break, we restricted the two series to the years 1920-1999. With these data, in all varieties of the ADF and PP tests with AR=1 or AR=2 the $P$-values are above 0.15 and the series thus modified can be accepted as I(1). Transforming ln GDP per capita and ln LEB into first differences, the stationarity tests in all its three varieties yield $P$-values below 0.001 so that the unit root null hypothesis is rejected at the usual levels of confidence and the conclusion is that at least the series including 1920-1999 data can be properly modeled as I(1) processes that become I(0), stationary, when transformed into first differences.

Applying the Johansen's cointegration test to these 80-year series, the test $P$-value for trace is below 0.0001 either for AR=$p$=1 or not specifying $p$ (in which case the SAS output indicates that since both the AR and MA orders are not specified, the test has been performed with AR=2 and MA=5 orders, which are determined by minimizing the information criterium). Because the null hypothesis of the Johansen's test is no cointegration, a *P*-value < 0.0001 implies to accept cointegration with a confidence level over 99.99%.

Chowdhury et al. claim there is a break in trend of LEB in 1917 and for that reason they compute all their statistics for two periods, 1841-1917 and 1918-1999. Because they do not provide sufficient detail for their analyses, we could not reproduce them. However, applying the Johansen cointegration rank test, we also arrive to the conclusion of cointegration, with the caveat that this is for the 1920-1999 data, because for 1841-1999 both the ADF and the PP tests reject the unit root null hypothesis. Thus, it seems we have been mostly able to reproduce, at least indirectly, the results of Chowdhury et al. We agree with Chowdhury et al. that income in the United Kingdom, proxied by ln GDP per capita, and population health in England and Wales, proxied by ln LEB, show cointegration when suitably transformed, according to widely used statistical tests.

From this result, the standard interpretation of cointegration (as discussed in Section 2) allows the conclusion that the two variables have some kind of "deep link" or causal connection in the long run. Chowdhury et al. estimate that the series are connected by a β long-run cointegration parameter which is approximately -0.4 for 1841-1917, and -0.8 for 1918-1999 (Tables 3 and 5 in Chowdhury et al.). They interpret these β estimates as "overwhelming evidence (…) in support of a negative relationship between life expectancy and income," but they do not explain the meaning of these numbers. Chowdhury et al. neither explain whether these negative 0.4 and 0.8 figures represent percentage points, years per thousand dollars, or any other unit that we may wonder is involved in this "negative relationship" between LEB and income they claim. Thus, Chowdhury et al. follow the standard interpretation of cointegration, by which their statistical tests are compelling evidence for a causal relationship between GDP per capita and LEB. However, is this causal connection plausible? Establishing causality is always difficult. Could this result be the consequence of faulty reasoning? Before answering that question, we will examine the analysis of this issue by Cavicchioli and Pistoresi with data from Italy.

Cavichioli and Pistoresi (2020) examined mortality rates, total, sex-specific, and some components of mortality for the years 1862-2013, and concluded that they are cointegrated with GDP per capita. These authors infer therefore that there is an equilibrium between these variables, so that, as shown in a previous quotation, an increase in real GDP per capita induces a reduction in mortality. The claim of a causal link is obvious and is accompanied by the assertion that their analysis is free from the possibility of omitted variables.

We tried to reproduce the results of Cavichioli and Pistoresi and could not get the mortality series that according to their paper were obtained from the National Institute of Statistics (ISTAT) of Italy. Instead, we computed total mortality, i.e., the crude death rate (CDR) for the years 1872–2013 from raw data on deaths and population in the Human Mortality Database. Because the CDR is a very poor indicator of population health, strongly modified by the age-structure of the population, we also obtained from the same source LEB data for Italy in the same years. We downloaded historical statistics of economic indicators of the Bank of Italy (all in nominal values), as well as data of the Italian real GDP per capita from the Maddison project. Figure 5 illustrates how a major decline in the Italian CDR and a substantial increase in LEB took place between 1870 and the 1950s, while in the most recent decades LEB continued growing but the CDR has remained basically flat—obviously, because of population aging.

The stationarity tests applied to the logarithm of the Italian CDR for the years 1872-2013 produce unclear results, with discrepancies between the tests. When computing the ADF test in SAS without specifying the AR and MA orders of the model (the SAS output indicates that AR =1 and MA=0 are selected by a minimum information criterion), for the zero-mean test, $P = 0.071$, so the null hypothesis of a unit root can be rejected at a 90% confidence level; the single mean and trend types of the test yield *P*-values above 0.50, thus the null is not rejected. In the PP test using 1 lag, the zero-mean test renders $P = 0.091$. Thus, we are rather unsure whether we should accept with Cavichioli and Pistoresi that the series of Italian log CDR must be considered as I(1).

Log real GDP per capita for the years 1872-2013 yields high *P*-values in the three types of the ADF and PP tests so the null that it is I(1) remains. The series of log LEB in the same years yields *P*-values above 0.39 in the zero mean and in the single mean types of the ADF and the PP tests, but in the trend type of the tests *P* = 0.017 (with AR=2 and MA=2 chosen by SAS), so the conclusion that the LEB series has a unit root is not well supported, despite the obvious rising trend of LEB in the period under consideration (Figure 5). At any rate, assuming that all these three series are I(1), we computed the Johansen rank cointegration test for ln GDP per capita with ln CDR and with ln LEB. In both cases we obtained a *P*-value below 0.001 when we set AR= *p* = 1, which seems the most parsimonious option. However, when we did the test without specifying the AR order, the output of the SAS program indicated that AR = MA = 2 had been chosen by a minimum information criterion; in this case, for cointegration of ln GDP per capita with LEB, *P* = 0.083; for cointegration of ln GDP per capita with ln CDR, *P* = 0.082. These results are statistically significant at the 90% level of confidence, but not at the 95% level. At any rate, since in economic research a 0.1 level of significance, that is, a 90% level of confidence is the usual threshold, we concluded (with minor caveats) with Cavichioli and Pistoresi that, for Italy, mortality (CDR) and GDP per capita are cointegrated (in logs). To that result, we add the result that GDP per capita is cointegrated with LEB, so that, we could say, the two indicators of population health, a very poor one (crude mortality, the CDR) and a good one (LEB), are cointegrated with the main indicator of income, GDP per capita.

Given that the notion of cointegration seems to us as rather unhelpful to understand the relationship between income and health, we tested whether the health indicators for Italy or for England and Wales appear also cointegrated in the Johansen test with other variables that according to the ADF and the PP tests are I(1).

## 5. Spurious cointegration

Using data from the ISTAT dataset of Italian historical statistics for periods of between 100 and 150 years we found Italian LEB cointegrated (1872-2018) with real GDP (not per capita), as well as with the nominal value of imports (1890-2015), with the value added (VA) in agriculture, with the VA in industrial activity, with the VA in services, and with the net industrial taxes (all these cointegrations tested with the Johansen test for the years 1872-2015). For annual data 1872-2020, Italian LEB appears also cointegrated with female, male and total births in Iceland (Figure 6). We also found that for the years 1957–2017, Italian LEB is cointegrated with the mean annual concentrations of $CO_2$ measured in the volcano Mauna Loa (the most used measure of atmospheric $CO_2$ concentrations); and for the years 1861-1947, Italian LEB is cointegrated with the annual series of the monetary aggregate M2 for Brazil.

We created three fake variables by transforming daily data 1997-1998 of Amazon stock in Wall Street into "annual" data starting in 1860. Using these fake annual series, according to the Johansen test Italian LEB is cointegrated (Figure 7) with both the open and the close value of the Amazon stock (both reveal a rising trend in the period) but not with the volume of Amazon stock transactions (which looks rather a stationary series and rendered *P*-values below 0.05 in the ADF test).

We tested whether health in England and Wales, as measured by ln LEB, appears cointegrated with other variables that according to the ADF test are I(1). We found that the series of ln LEB for England and Wales appears cointegrated with GDP per capita for Spain (in international Geary-Kamis dollars of 2011), as well as with the average annual price of oil in international markets in current dollars, and with two US series of quarterly data that we "transformed" into fake annual data, the consumer price index (CPI) and the average wage in dollars per hour of manufacturing workers.

We also tested whether using the Johansen test we could find cointegration of LEB series for England and Wales or Italy with computer-generated series (Figure 8). We generated random walks $y_t = y_{t-1} + \varepsilon_t$, where $\varepsilon_t$ is randomly distributed with mean μ. We tried many series and found cointegration of one of these series with LEB or mortality in a large majority of the trials when μ ≠ 0, that is, when the generated series is a random walk *with* drift.

Having found so many different I(1) series cointegrated with a major indicator of population health as LEB, we thought it would be interesting to explore whether it is possible to find cointegration between series that *a priori* look totally unrelated. Using the Italian historical statistics provided by the ISTAT, we found total VA cointegrated with the value of imports, which in turn is cointegrated with the VA in services. Also, the VA in construction is cointegrated with the value of imports, the VA in industry is cointegrated with the VA in construction, and with the VA in services. We also found the fake Amazon series of open daily quotes of Amazon stock cointegrated with the VA in Italian industry in 1860-2017, the VA in Italian agriculture, Italian GDP at market prices, Italian real GDP in dollars of 2011, and Italian population. Random walks with drift generated with SAS as formerly explained were found to be cointegrated with the VA in agriculture, nominal GDP, value of imports and other components of the Italian national accounts. They were also cointegrated with a 1872-1997 series of Swedish nominal GDP. We also found cointegration between Swedish and Italian nominal GDP. We found total annual births in Iceland in 1850-1998 cointegrated with the Swedish nominal GDP. The annual mortality rate (CDR) of the US is cointegrated with AR = $p$ = 2 with the number of registered firms in business for the period 1901-1997 (Figure 9). We also found that for annual data 1875-2004, New Zealand's LEB is cointegrated with the population of Sweden. Because the cointegration equation can include a negative cointegrating parameter, it is perfectly possible to find cointegration between a growing series and a declining series, as illustrated by Figures 8 and 9, or by the notion in the papers by Brenner (2005) or Cavicchioli and Pistoresi (2020) that GDP per capita, obviously rising in the long run, is cointegrated with mortality, which is a declining series.

Spurious claims arising from standard tests could simply be a consequence of widespread misuse of the tests. Perhaps assumptions have to be checked more carefully, or more sophisticated tests applied? We argue that the issue is more fundamental, primarily due to the difficulty of distinguishing stochastic from non-stochastic trends. This has been noticed previously, but our investigation of cointegration between LEB and GDP per capita shows that lessons remain to be learned.

## 6. Discussion

Since the notion of cointegration was introduced in econometrics about half a century ago, spurious results of cointegration tests have been mentioned sometimes in the econometric literature. Franses (1990) referred directly to spurious cointegration in the title of his research note. Cheung & Lai (1993) noticed that Johansen's tests are seriously biased toward finding spurious cointegration. Wickens (1996) warned that the existence of common stochastic trends is difficult to test and usually cannot be established without introducing restrictions that are not part of tests for cointegration, which probably makes cointegration much less valuable than was first expected. Gonzalo and Lee (1998) showed that, in quite common situations, Johansen tests "tend to find spurious cointegration with probability approaching one asymptotically". More recently, Hjalmarsson and Österholm (2007a, 2007b, 2010) have shown that there is a substantial probability of falsely concluding that completely unrelated trending series are cointegrated.

These considerations on the pitfalls of cointegration, however, seem to have been mostly overlooked in econometric theory and practice, as the notion of spurious cointegration is overwhelmingly ignored. Indeed, in standard econometric textbooks like those by Baltagi (2008), Brooks (2008), Davidson and MacKinnon (1999), Lütkepohl (2005), or Stock and Watson (2019), the notion of spurious cointegration is not even mentioned despite the fact that spurious regression is discussed in detail. In the "cointegration" entry of Wikipedia (checked in January 2025) there is no mention of the possibility that the cointegration can be spurious; there is neither an entry for "spurious cointegration". In books of applied econometrics using R (Kleiber & Zeileis 2008), or Stata (Baum 2006), or SAS (Ajmani 2009), there is no mention nor discussion of spurious cointegration.

Cointegration methods were developed by and for econometric data analysis. They are currently used rarely outside econometric contexts. Yet, econometrics is a major discipline in the quantitative social sciences, and our case study shows that the proper limitations of cointegration are not well known.

Cointegration testing starts with stationarity (unit root) tests of which a variety exist. Each of these tests yields a statistic with an associated *P*-value. Supposedly the test itself needs to be expertly tuned to the characteristics of the particular time series at hand by choosing the proper AR and MA orders of the testing model, or one has to extend the test to include automatic selection of these variables. Because of all this, the interpretation of the stationarity tests is far from being straightforward. Despite warning like those by Wickens (1996) or Gonzalo and Lee (1998), papers asserting the presence or absence of cointegration among various time series, very often ignore these issues. In the view of Hjalmarsson and Österholm (2007a), for many time series "it is difficult to justify theoretically that they are generated by unit root processes." From there, a faulty demonstration of cointegration follows.

For unit root tests, neither the null hypothesis nor the alternative hypothesis considers nonstationary models with general nonlinear (but non-random) trend. Therefore, unit root testing is inappropriate to distinguish models with nonlinear deterministic trends from those with stochastic trends. Since only the latter are generally I(1), any conclusion that a time series should be modeled as I(1) is based on an assumption that a unit root test is unable to validate. Therefore, it becomes a matter of belief for the data analyst whether to construct a model with a unit root (in which case one can talk of cointegration) or a nonstationary model where there is not a unit root (in which case, one cannot). Assuming that observed time series generally have stochastic trends (Stock and Watson 2019:541) primes up the analysis toward finding cointegration.

The use of cointegration tests to argue for causal relationships between different components of a large and complex social system is problematic so far as the evidence for cointegration is driven by low frequencies. Having series of many data points, say many hundreds or thousands of observations (e.g., daily observations for a period of many years), there might be sufficient evidence in medium frequencies to suggest that variables tend to "move together" more reliably than could be expected by chance. At any rate, low frequency relationships are well known to be a poor basis for causal reasoning, particularly with time series with a few hundred data points each, as is common in social science research. Via its Fourier decomposition, a time series can be explained as a sum of oscillations of different frequencies. If two series are detrended by one of the available means (differencing, subtraction of a moving average, whatever), thus eliminating low-frequency components of it, and the residuals of the two series after this detrending reveal a high correlation, this is giving us strong evidence, evidence still based on hundreds of data points, suggesting some causal link between these series or between these two series and a third one. Contrarily, cointegration is based in finding a common "stochastic" trend, that is, some commonality in the low-frequency component of the series. By abstracting a trend from a series, we are just discarding most of the information in the data that form the series. Since, as shown by Hjalmarsson and Österholm, to ascertain whether a trend is or is not stochastic is rather difficult, it seems to us that most applications of cointegration just check for common trends. But there are few possible potential trends in a series, so that it is easy that just by chance a long time series matches the trend of a completely unrelated variable. Thus, the real issues in the elucidation of causality can be obfuscated by dressing up a search for low-frequency relationships in the language of cointegration.

In recent years causal analysis and statistical practice has been subjected to major criticisms and reconsiderations (for instance in Pearl & Mackenzie 2018, Mayo 2018; Clayton 2022). The notion of spurious cointegration was presented long time ago but it has been largely ignored, despite the fact that, as we have shown, given a series with a trend and a unit root

"demonstrated" by a proper test, it is quite easy to find other series with unit roots that in the Johansen test appear cointegrated with the former.

We conclude that there is not plausible evidence that GDP per capita and LEB have any "equilibrium" in the long run because its cointegration is just a spurious one. More generally, we conclude that the term "spurious cointegration," presently *rara avis* in the econometric literature and in applied research, should be a proper companion to "spurious regression". "Correlation does not imply causation" is an accepted rule in social science. After finding many cases of a series that appears cointegrated with totally unrelated ones, it seems to us that "cointegration does not imply causation" is also a very proper rule.

**APPENDIX**

We employed SAS for most of the statistical procedures, using the macro %DFTEST and the procedure PROC ARIMA to compute the ADF and the PP tests. We used PROC VARMAX to compute the Johnasen's test of cointegration. For instance, the code

```
proc arima;
identify var=ln_e0 stationarity=(adf=1); run;
identify var=ln_e0 stationarity=(pp=1); run;
```

yields the results of the ADF and PP stationarity tests for the variable ln_e0 performed with a model without lag of with one lag only. The code

```
%dftest (A001, ln_e0, ar=0, trend=0, outstat=T1 );
```

yields the *P*-value of the ADF test for ln_e0 computed with the A001 dataset and a no lag (ar=0) and zero mean (trend=0) model. If trend=1 or trend = 2 is specified, the results are respectively for the single mean and trend varieties of the ADF test.

The following code

```
proc varmax; model ln_Ypc ln_e0 / p = 1 noint cointtest=(johansen);
```

yields the trace for the Johansen's test of cointegration and the corresponding *P*-value using an ARIMA model with AR = $p$ = 1. If "p = 1" is suppressed, SAS produces a note that explains that "[s]ince both the autoregressive and moving average orders are not specified," the AR and MA orders of the ARIMA model used for the test are selected by a minimum information criterion.

We reproduced some of the results obtained with SAS with the program *R*, using the ur.df function.

**Figure 1**. LEB (years) for Sweden, the Netherlands, and England and Wales. Note the health crises in Sweden, for instance in 1772 and 1809, and in the three series in 1918-1919, because the world flu pandemic. The impact of World War II in 1940-1945 in noticeable in the two locations affected by the war.

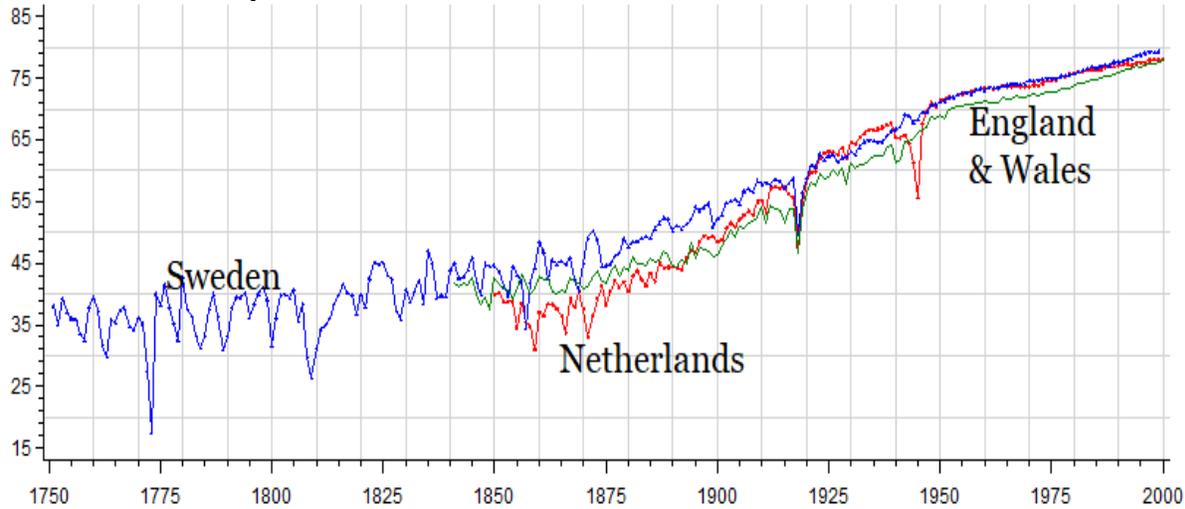

Data for period LEB from the Human Mortality Database. The England and Wales series corresponds to the civilian population.

**Figure 2**. Life expectancy at birth (years) in nine countries, 1990-2021

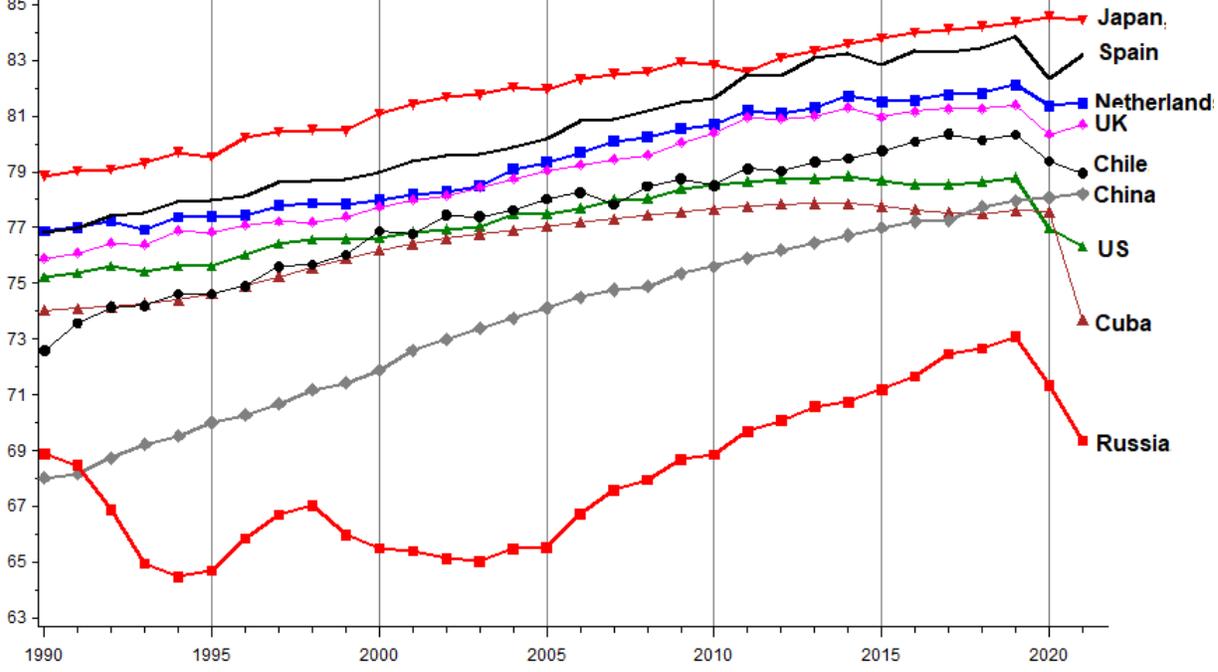

Data from the World Development Indicators database of the World Bank, downloaded March 2024

**Figure 3**. Life expectancy at birth for the civilian (LEBc) and the total (LEBt) population of England and Wales (years, l. h. s.) and GDP per capita (GDPpc) for the United Kingdom, in international Geary-Kamis dollars of 1990

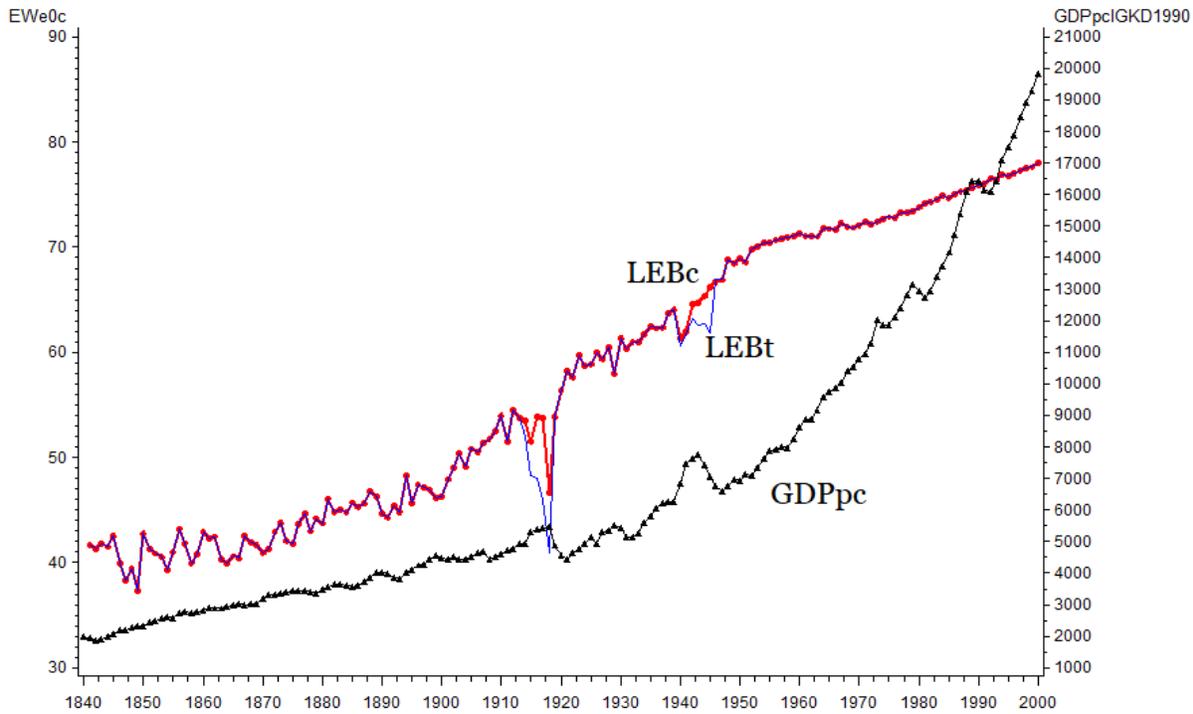

LEB data from the Human Mortality database, GDP per capita from Maddison Project.

**Figure 4.** United Kingdom GDP per capita (GDPpc, 1990 international Geary-Kamis dollars, l.h.s.) and life expectancy at birth (LEB) for the civilian population of England and Wales (years, r.h.s.), both series in natural logs. The upper and the lower panels only differ in the right-hand scale for ln LEB.

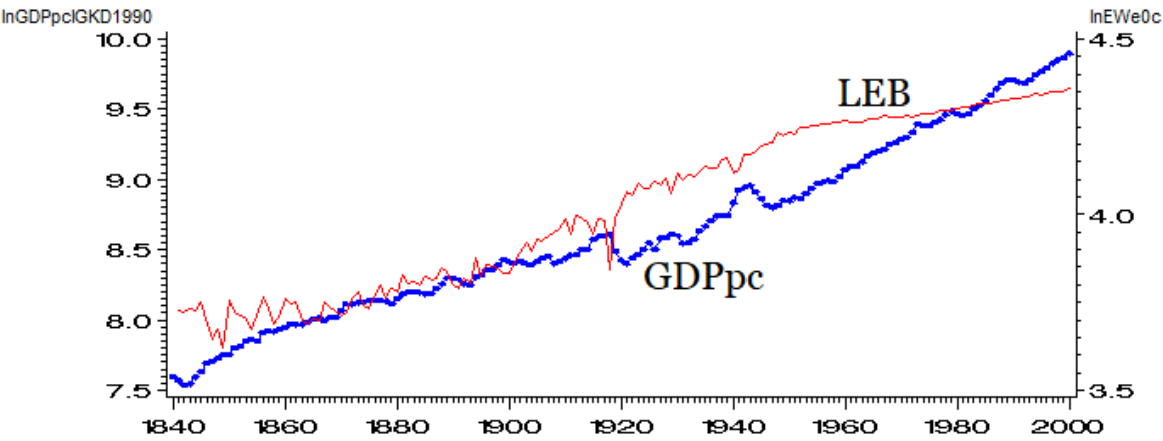
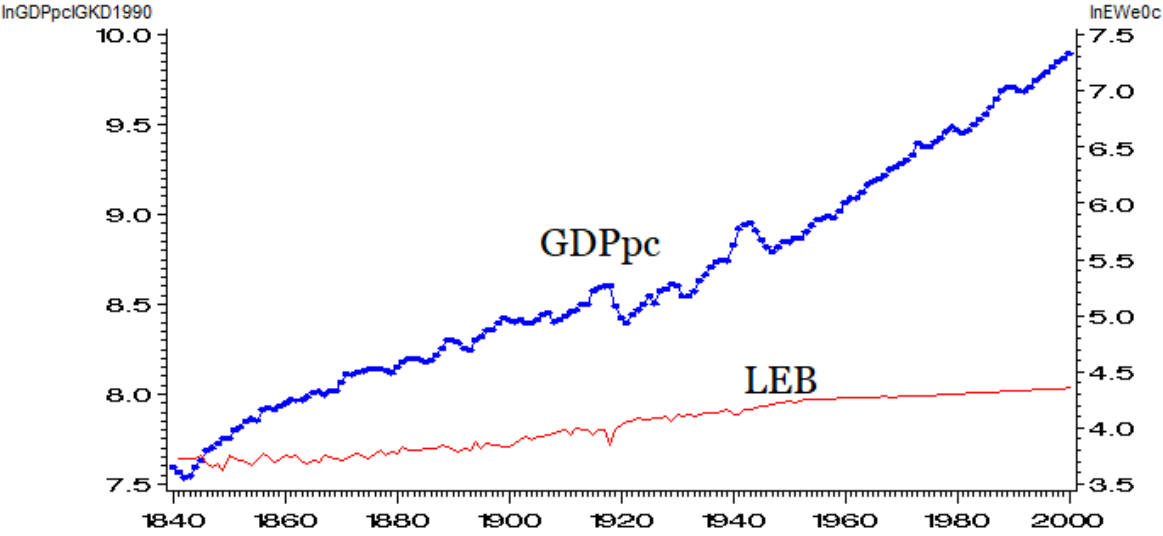

**Figure 5**. Crude mortality rate (CDR, deaths per thousand, l.h.s.), life expectancy at birth (LEB, years, l.h.s.) and GDP per capita (r.h.s., log scale), Italy 1870-2013. Note the major effects of the world flu pandemic in 1918-1919 and World War II in 1939-1945 on LEB and the CDR. The Great Recession of 2009 is obvious in the GDP per capita curve but not in the other curves.

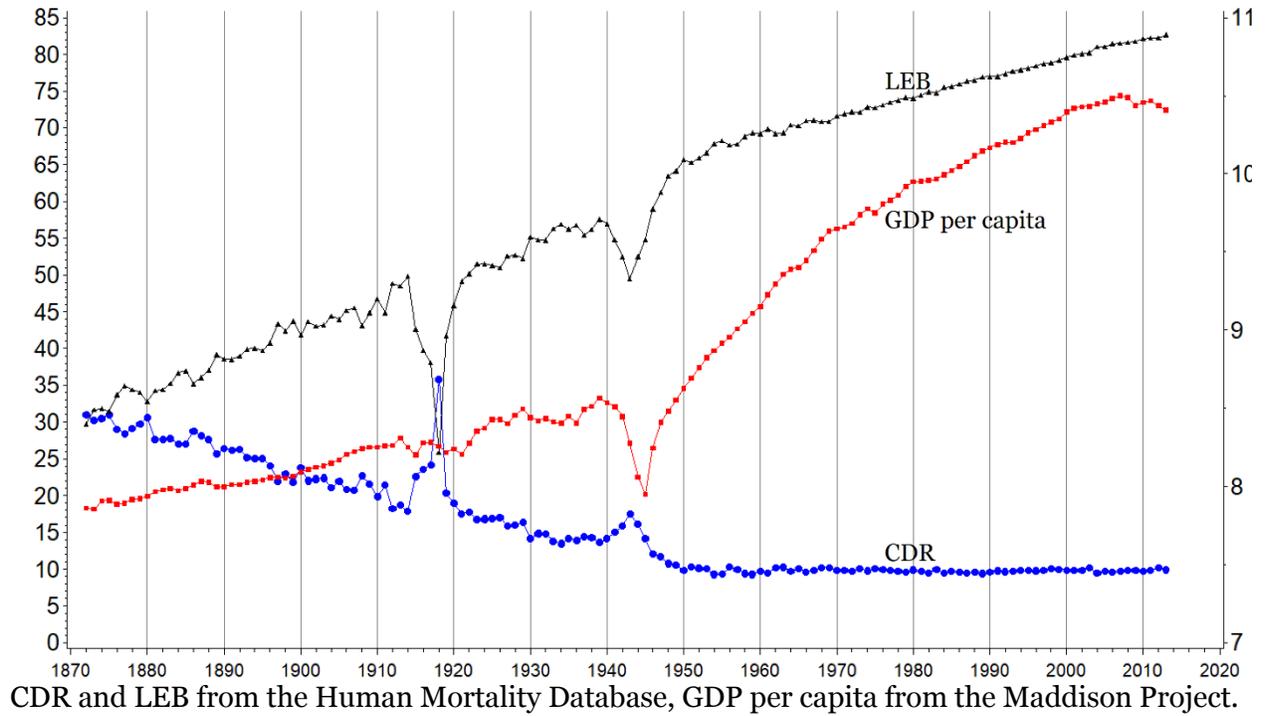

CDR and LEB from the Human Mortality Database, GDP per capita from the Maddison Project.

**Figure 6**. Life expectancy at birth in Italy (dots, years, r.h.s.) and annual births in Iceland (triangles, l.h.s.) in the past 160 years. In the Johansen trace test applied to these two series for the period 1872-2020, the null of no cointegration is rejected ($P < 0.001$).

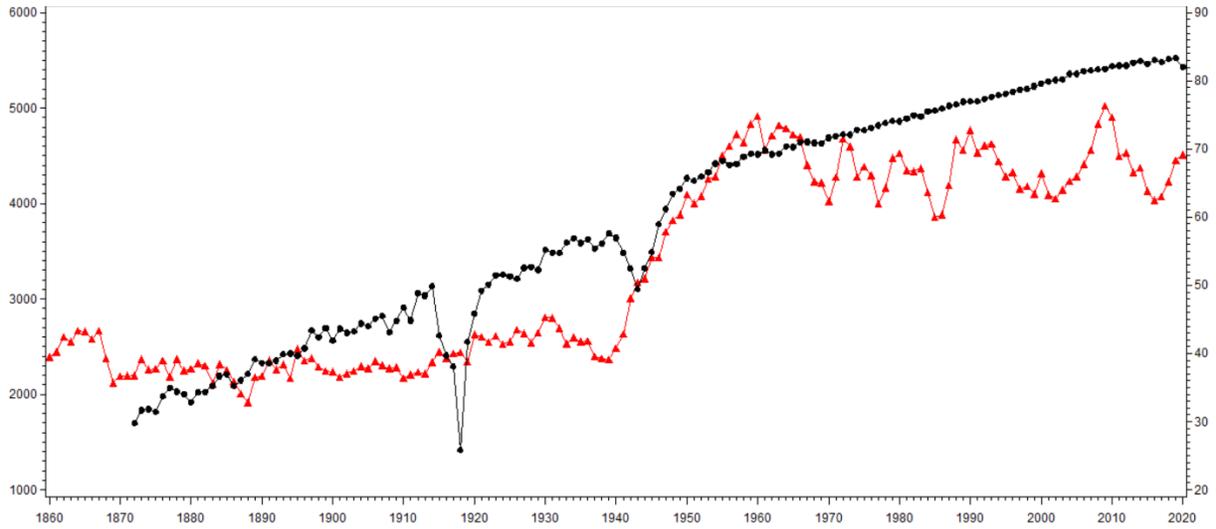

**Figure 7**. Life expectancy at birth (LEB, years, r.h.s.) in Italy (1872-2020) and a fake annual series (triangles) obtained from making annual the daily open values of the Amazon stock in 1997-1998. The two series appear cointegrated in the Johansen test.

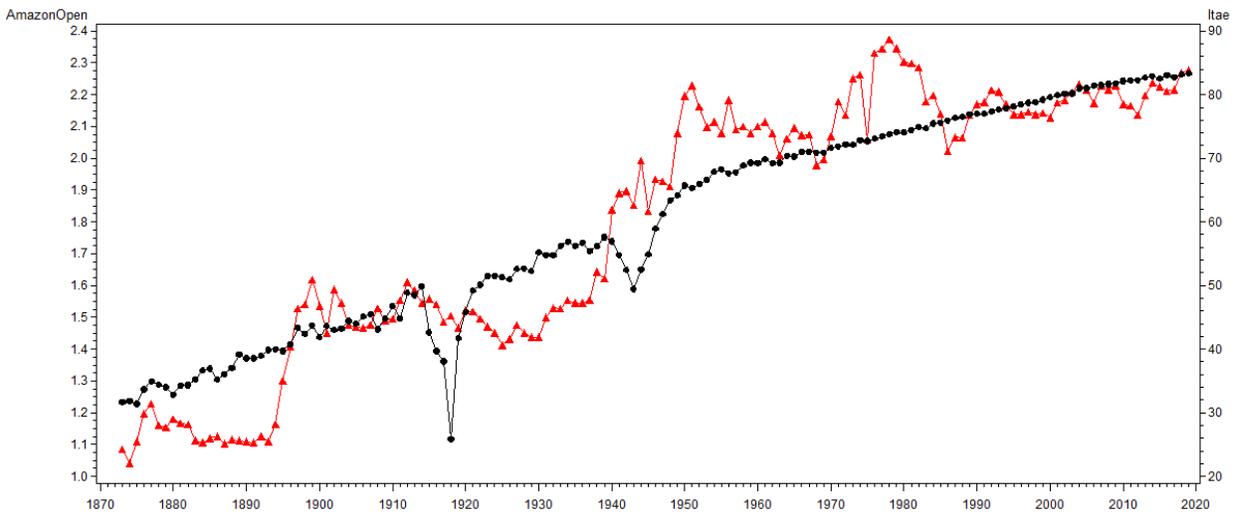

**Figure 8**. Three random walks (l.h.s.) that according to the Johansen test are cointegrated with England & Wales's LEB (in natural logs, thick line, r.h.s.).

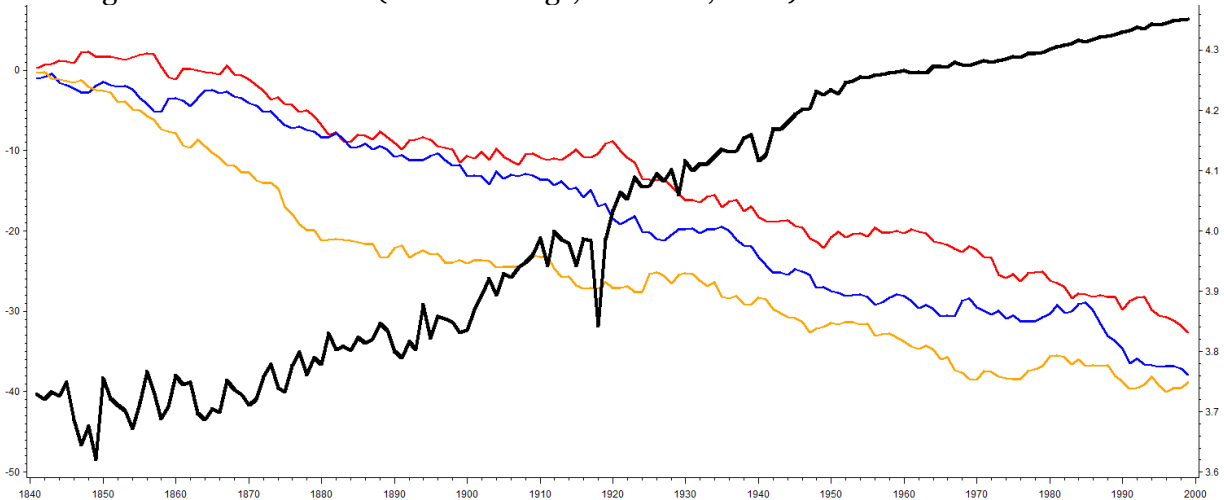

The random walks are three realizations of the process $y_t = y_{t-1} + \varepsilon_t$, in which $\varepsilon_t$ is a randomly distributed Gaussian error with mean $\mu = -0.2$ and standard deviation $s = 0.7$.

**Figure 9**. Crude mortality rate (deaths per thousand population) and number of active firms in business in the United States. According to the Johansen test the two series are cointegrated.

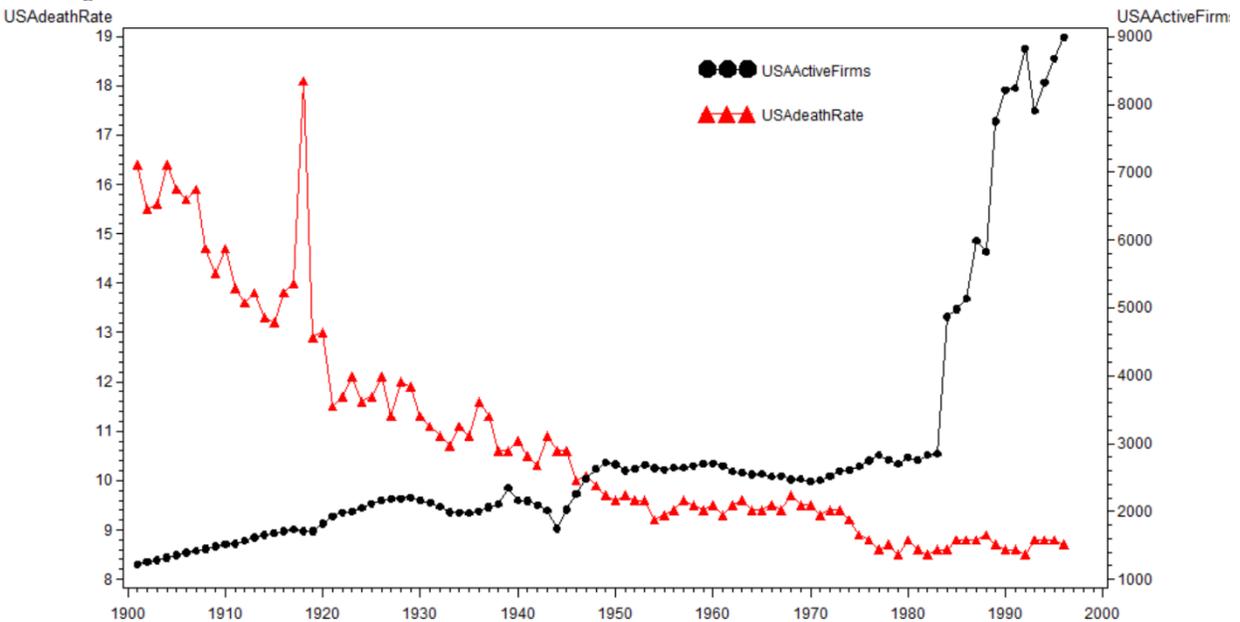